\documentclass[journal]{IEEEtran}
\usepackage[T1]{fontenc}
\usepackage[latin9]{inputenc} \usepackage{color} \usepackage{refstyle}
\usepackage{float} \usepackage{calc} \usepackage{amsmath}
\usepackage{amssymb} \usepackage{subcaption} \usepackage{graphicx}

\usepackage{soul}
\usepackage{comment}

\usepackage[unicode=true,
bookmarks=true,bookmarksnumbered=true,bookmarksopen=true,bookmarksopenlevel=3,
breaklinks=true,pdfborder={0 0
  0},pdfborderstyle={},backref=section,colorlinks=true]{hyperref}
\hypersetup{pdftitle={Your Title}, pdfauthor={Your Name},
  pdfpagelayout=OneColumn, pdfnewwindow=true, pdfstartview=XYZ,
  plainpages=false}

\makeatletter
\begin{document}
\hypersetup{%
    ,urlcolor=black
    ,citecolor=green
    ,linkcolor=blue
    }
\font\myfont=cmr12 at 18pt
\title{\myfont Uniqueness Theorem: With Normal Components  Specified on  External Spherical Surface}

\author{
    \IEEEauthorblockN{Rajavardhan Talashila\IEEEauthorrefmark{1},}
    \\
    \IEEEauthorblockA{\IEEEauthorrefmark{1} University of Colorado Boulder
    \\ } 
    }
%\markboth{Submitted to  IEEE Antennas and Wireless Propagation Letters}{Your Name \MakeLowercase{\emph{et al.}}: Your Title}
\maketitle
\begin{abstract}
  A  uniqueness theorem for time-harmonic electromagnetic fields which requires the
  normal components of electromagnetic fields specified on a spherical surface is proposed and  proved. The
  statement of the theorem is : "For a
  spherical volume $V$ that contains only perfect conductors and
  homogeneous lossless materials and for which the
impressed currents $\mathbf{J}$ are specified, a time-harmonic solution
to the Maxwell's equations within the volume, having  outgoing
waves alone,  is uniquely specified by the values of the
radial components of both $\mathbf{E}$ and $\mathbf{B}$ over the
exterior spherical
surface $V$ and the tangential components of either $\mathbf{E}$ or
$\mathbf{B}$ on the interior surfaces."  The proof of this theorem
relies on  the uniqueness of multipole
expansion of electromagnetic fields outside the enclosing sphere. The
conventional uniqueness theorem for the  volume $V$ having
loss-less materials is considered to be
the case of lossy materials in the limit the dissipation approaching zero.
\end{abstract}

\section{Introduction}

\IEEEPARstart{U}{niqueness} theorem in classical electromagnetism
allows to know what information is required for finding the solution
and makes sure that the obtained solution will be unique. As stated by
\cite{harrington_time-harmonic_2001}, it allows to establish a
one-to-one correspondence between the sources and the fields for a
given set of boundary conditions.

The conventional uniqueness theorem for the  electric ($\mathbf{E}$)
and magnetic ($\mathbf{B}$) fields inside a volume needs the  tangential components of the $\mathbf{E}$ or $\mathbf{B}$
fields or both  specified on the surface the enclosing volume. In this
work,  we prove that for  uniqueness of the electromagnetic fields inside a volume,  enclosing all the sources, having
the outer surface as a sphere,  it is sufficient to specify  the
radial components of both  $\mathbf{E}$ and $\mathbf{B}$
fields on the outer spherical surface and the tangential components on
the inner surface. The region outside the considered spherical volume
is assumed to be completely homogeneous which translates to having
only  outgoing waves from the sources and no waves entering into he
region. This is precisely the situation for most of the antenna
simulations where the antenna is bounded in a box of homogeneous material.

The equivalence of specifying the radial components of the
$\mathbf{E}$ and $\mathbf{B}$ fields and the  tangential
components of either $\mathbf{E}$ or $\mathbf{B}$ on the outer sphere
is achieved by  using the multipole expansion of the fields outside the
sphere. The multipole expansion of the fields outside a source free
region is unique. The uniqueness in terms of the radial components of
$\mathbf{E}$ an $\mathbf{B}$ is proved in \cite{bouwkamp_multipole_1954}.
It will be shown, mathematically, that such uniqueness of the fields
of the multipole expansion also holds with
either the tangential $\mathbf{E}$ or $\mathbf{B}$ fields.
Though multipole expansion is well studied and applied for various
cases \cite{jackson1999classical}, the above said equivalence is not brought out.

This short communication has the conventional uniqueness theorem stated in section
\ref{sec:Uniqueness-conventional}, and the mathematical details of the
multipole expansion in section \ref{sec:multipole} demonstrating the
equivalence of specifying the tangential and radial components on a
spherical surface. The  proposed theorem and its logical proof is presented in  section \ref{sec:Proof}.

% Journal Papers on Uniqueness Theorem

% Feynman on Uniqueness theorem

% Taflove, Maxwell, Stratton, Jackson on Uniqueness theorem

% Taflove

% Note:
% 1: Charge density is in accordance with the continuity equation and is
% non-zero
% 2. Sources (currents and charges are inside the volume V)

\section{ Uniqueness Theorem}
\label{sec:Uniqueness-conventional}
The uniqueness theorem for the time-harmonic fields as
stated by  \cite{smith_introduction_1997} is:

% Glenn S. Smith
"For a Volume V that contains only simple-materials and for which the
impressed currents \textbf{J} are specified, a time-harmonic solution
to the Maxwell's equations is uniquely specified by the values of the
tangential components of either \textbf{E} or \textbf{B} over the
internal and external boundary
surfaces of $V$."

Here simple materials mean linear, isotropic and homogeneous materials. The
proof of the above theorem, which is based on the complex Poynting's
theorem,  assumes the 
volume to have lossy materials, with the exception of
PEC conductors
\cite{smith_introduction_1997}\cite{balanis_advanced_2012}\cite{stratton_electromagnetic_2012}.
And the case of volume $V$ with lossless materials ($\sigma =0$), the
fields can be considered to be the limit of the corresponding fields
for a lossy medium as the dissipation approaching zero.

\section{ Multipole Theory}
\label{sec:multipole}
The general solution of the fields in source free region, with
$e^{-j\omega t}$ time dependence, to the Maxwell's equations in
multipole expansion 
form considering  outgoing waves alone is  \cite{jackson1999classical}\cite{bouwkamp_multipole_1954}\cite{papas1988theoryof}

\begin{align}
  \mathbf{E} & =
               Z_{0}\sum_{l,m}\Bigg[\cfrac{i}{k}a_{E}(l,m)\nabla\times\left(h^{(1)}_{l}\left(kr\right)\boldsymbol{X}_{lm}\right)\nonumber\\
  &\qquad \qquad \qquad +a_{M}(l,m)h^{(1)}_{l}\left(kr\right)\boldsymbol{X}_{lm}\Bigg]\label{eq:E-Jackson}\\
  \mathbf{B} &
               =\mu_{0}\sum_{l,m}\Bigg[a_{E}(l,m)h^{(1)}_{l}\left(kr\right)\boldsymbol{X}_{lm}
               \nonumber \\
                &\qquad \qquad \qquad -\cfrac{i}{k}a_{M}(l,m)\nabla\times\left(h^{(1)}_{l}\left(kr\right)\boldsymbol{X}_{lm}\right)\Bigg]\label{eq:H-Jackson}
\end{align}
where
\begin{align}
\mathbf{X}_{lm}\left(\theta,\phi\right) &
                                          =\cfrac{1}{\sqrt{l\left(l+1\right)}}\cfrac{1}{i}\left(\boldsymbol{r}\times\nabla\right)\left(Y_{lm}\left(\theta,\phi\right)\right)
 \nonumber \\ 
 & =\cfrac{1}{\sqrt{l\left(l+1\right)}}\cfrac{1}{i}\left(-
   \cfrac{\partial
   Y_{lm}}{\partial\theta}\hat{\phi}+\cfrac{1}{\sin\theta}\cfrac{\partial
   Y_{lm}}{\partial\phi}\hat{\theta}\right) \label{eq:Xlm}
\end{align}
The multipole expansion of the  electric field expressed with the
longitudinal ($\hat{r}$) and transverse components ($\hat{\theta},
\hat{\phi}$) separately is
\begin{align}
  \mathbf{E} & = Z_{0}\sum_{lm}^{}\Bigg[ a_{E}(l,m)
                              \cfrac{1}{kr} \sqrt{l(l+1)} h^{(1)}_l(kr)
                              \mathbf{Y}_{lm} \nonumber \\
  &\qquad \qquad + a_{E}(l,m) \cfrac{i}{kr}\cfrac{\partial}{\partial r}
    (rh^{(1)}_{l}(kr)) \mathbf{Z}_{lm} \nonumber
\\
  &\qquad
    \qquad+a_{M}(l,m)h^{(1)}_{l}\left(kr\right)\boldsymbol{X}_{lm}\Bigg]
  \label{eq:E-expanded}\\
    \mathbf{B} & = \mu_{0}\sum_{lm}^{}\Bigg[ a_{M}(l,m)
                              \cfrac{-1}{kr} \sqrt{l(l+1)} h^{(1)}_l(kr)
                 \mathbf{Y}_{lm}\nonumber \\
  &\qquad
    \qquad+a_{E}(l,m)h^{(1)}_{l}\left(kr\right)\boldsymbol{X}_{lm}
    \nonumber \\
  &\qquad \qquad - a_{M}(l,m) \cfrac{i}{kr}\cfrac{\partial}{\partial r}
    (rh^{(1)}_{l}(kr)) \mathbf{Z}_{lm} \Bigg]\label{eq:H-expanded}
\end{align}

where 

\begin{align}
   \mathbf{Z}_{lm}(\theta,\phi) &= \hat{r} \times
                                  \mathbf{X}_{lm}\left(\theta,\phi\right)
  \nonumber \\
    &=\cfrac{1}{\sqrt{l\left(l+1\right)}}\cfrac{1}{i}\left(\cfrac{\partial
    Y_{lm}}{\partial\theta}\hat{\theta}+\cfrac{1}{\sin\theta}\cfrac{\partial
    Y_{lm}}{\partial\phi}\hat{\phi}\right) \label{eq:Zlm}
\end{align}
and
\begin{equation}
  \label{eq:Ylm}
\mathbf{Y}_{lm} = Y_{lm}\hat{r}
\end{equation}

where  $Y_{lm}$ is the spherical harmonic function of degree $l\ge0$ and order $m\,\left(-l\le m\le l\right)$ defined
as
\begin{equation}
  Y_{lm}\left(\theta,\phi\right)=\sqrt{
    \cfrac{(2l+1)}{4\pi}\cfrac{\left(l-m\right)!}{\left(l+m\right)!}}P_{l}^{m}\left(\cos\theta\right)e^{im\phi} \label{eq:Ylm-sph-harmonic}
\end{equation}

The orthonormal conditions of the Spherical vector harmonic functions
$\mathbf{Y}_{lm}$, $\mathbf{X}_{lm}$ and $\mathbf{Z}_{lm}$ are

\begin{subequations}
\begin{align}
  \int \mathbf{X}_{l'm'} \cdot \mathbf{X}^{*}_{lm} d\Omega &= \delta_{lm} \delta_{l'm'}\\
  \int \mathbf{Z}_{l'm'} \cdot \mathbf{Z}^{*}_{lm} d\Omega &= \delta_{lm} \delta_{l'm'}\\
  \int \mathbf{Y}_{l'm'} \cdot \mathbf{Y}^{*}_{lm} d\Omega &= \delta_{lm} \delta_{l'm'}\\
  \int \mathbf{X}_{l'm'} \cdot \mathbf{Z}^{*}_{lm} d\Omega &= 0
  \\
  \int \mathbf{X}_{l'm'} \cdot \mathbf{Y}^{*}_{lm} d\Omega &= 0
  \\
  \int \mathbf{X}_{l'm'} \cdot \mathbf{Y}^{*}_{lm} d\Omega &= 0                    \end{align}  \label{eq:Orthonormal-Y-X-Z}
\end{subequations}

for all the possible values of $l$, $l'$, $m$ and $m'$.

Using the orthormal relations in eq.~(\ref{eq:Orthonormal-Y-X-Z}) and
the multipole expansion of the electric and magnetic fields in
eq.~(\ref{eq:E-expanded})~(\ref{eq:H-expanded}), the  multipole
coefficients of the expansion  can be obtained using only the radial
components of the electric and magnetic fields on a sphere of radius $r_0$ enclosing
the sources from:

\begin{subequations}
  \label{eq:Coeff_radial}
\begin{align}
   a_{E}\left(l,m\right) 
  &=\cfrac{1}{Z_{0}}\cfrac{1}{h_{l}^{\left(1\right)}\left(kr_0\right)}
    \cfrac{kr_{0}}{\sqrt{l(l+1)}}\int
    \boldsymbol{Y}^*_{lm}\boldsymbol{\cdot}\mathbf{E}d\Omega  \\
a_{M}\left(l,m\right)  
  &=\cfrac{-1}{h_{l}^{\left(1\right)}\left(kr_0\right)}
    \cfrac{kr_{0}}{\sqrt{l(l+1)}}\int
    \boldsymbol{Y}^*_{lm}\boldsymbol{\cdot}\mathbf{H}d\Omega  
\end{align}
\end{subequations}

Similarly, the multipole coefficients can be obtained  from the
tangential components of the electric field alone using:
\begin{subequations}
  \label{eq:Coeff_Electric}
\begin{align}
  a_{E}\left(l,m\right)  
  &=\cfrac{r_{0}}{Z_{0}} \left(\left[\cfrac{\partial}{\partial r}
    (rh^{(1)}_{l}(kr) \right]_{r=r_{0}} \right)^{-1}
    \int
  \boldsymbol{Z}^*_{lm} \boldsymbol{\cdot} \mathbf{E} d\Omega \\
  a_{M}\left(l,m\right)  
 &=\cfrac{1}{Z_{0}h_{l}^{\left(1\right)}\left(kr_0\right)}\int
    \boldsymbol{X}^*_{lm}\boldsymbol{\cdot}\mathbf{E}d\Omega  
\end{align}
\end{subequations}

and finally, the multipole coefficients can be obtained  from the
tangential components of the magnetic field alone using:
\begin{subequations}
  \label{eq:Coeff_Magnetic}
\begin{align}
  a_{E}\left(l,m\right)  
 &=\cfrac{1}{h_{l}^{\left(1\right)}(kr_{0})}\int
   \boldsymbol{X}^*_{lm}\boldsymbol{\cdot}\mathbf{H}d\Omega  \\
    a_{M}\left(l,m\right)  
  &=r_{0} \left( \left[ \cfrac{\partial}{\partial r} (rh^{(1)}_{l}(kr)\right]_{r=r_{0}} \right)^{-1}
    \int
  \boldsymbol{Z}^*_{lm} \boldsymbol{\cdot} \mathbf{H} d\Omega 
\end{align}
\end{subequations}
The set of equations
(\ref{eq:Coeff_radial})(\ref{eq:Coeff_Electric})(\ref{eq:Coeff_Magnetic})
lead to the conclusion that the fields outside the sphere containing the sources can be fully
characterized equivalently in the following ways:
\begin{itemize}
\item only the radial components of the electric and
magnetic fields on the sphere
\item only the tangential electric fields on the sphere
\item only the tangential magnetic fields on the sphere
\end{itemize}

By utilizing  this equivalent representation of the fields outside the
sphere  and the conventional uniqueness theorem stated in section
\ref{sec:Uniqueness-conventional}, a new theorem is proposed in the
following section.

\section{Proposed Theorem}
\label{sec:Proof}
\begin{figure}[ht]
  \centering
\includegraphics[width=6.5cm]{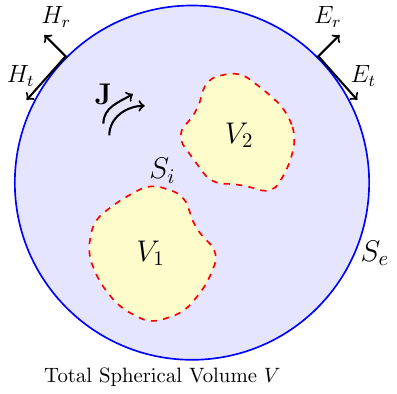}  
  \caption{\label{fig:Volume-description} The representation of the
    volume $V$ with external spherical surface $S_{e}$ and internal
    surface $S_{i}$. It has the source currents $\mathbf{J}$ specified. }
\end{figure}

The proposed theorem is:  "For a
  spherical volume $V$ that contains only perfect conductors and
  homogeneous lossless materials and for which the
impressed currents $\mathbf{J}$ are specified, a time-harmonic solution
to the Maxwell's equations within the volume, having radially outgoing
waves alone,  is uniquely specified by the values of the
radial components of both $\mathbf{E}$ and $\mathbf{B}$ over the
exterior spherical
surface $V$ and the tangential components of either $\mathbf{E}$ or
$\mathbf{B}$ on the interior surfaces."

The spherical volume as shown in
Fig.~\ref{fig:Volume-description} with a spherical external surface
$S_{e}$ and internal surfaces $S_{i}$ due to the internal volumes
$V_{1}$, $V_{2}$ etc. It has the impressed source currents $J$
specified. Now, Consider the case where the radial
components of the electromagnetic fields on the exterior surface
$S_{e}$ and the tangential electric or magnetic fields are specified
on the internal surface $S_{i}$. However, specifying the radial components of the $\mathbf{E}$ and $\mathbf{B}$
fields on the enclosing sphere is equivalent to specifying the
tangential components of the electric fields or magnetic fields on
it for the unique description of the fields outside the sphere
including the spherical surface.   

This amounts to specifying the
tangential components of  $\mathbf{E}$ or $\mathbf{B}$ on the
interior $S_i$ and exterior $S_{e}$ surfaces of volume $V$. Hence, 
the uniqueness of the fields inside the spherical volume $V$ is justified using the conventional uniqueness theorem presented in
sec~\ref{sec:Uniqueness-conventional}.

\section[conclusion]{Conclusion}
\label{sec:conclusion}
The multipole expansion of the electromagnetic fields outside a sphere
enclosing all the source in a homogeneous loss-less medium  is used to
show the equivalence of specifying  the tangential $\mathbf{E}$ or
$\mathbf{B}$ fields to that of specifying the radial components of both
the $\mathbf{E}$ and $\mathbf{B}$ fields. This is in turn used to show the
uniqueness of the fields inside the spherical volume enclosing the
source currents with the radial components of both $\mathbf{E}$ and
$\mathbf{B}$ fields specified on the spherical surface. This
uniqueness theorem can also be extended to arbitrary surface enclosing
the sources using the theorem proved by Rumsey in
\cite{rumsey_new_1959}.

\bibliographystyle{IEEEtran}
\bibliography{Uniqueness,F_on_sphere}

\section{Half-wave Dipole}

To validate the general procedure of calculating the far fields from the
radial fields on the surface of a sphere, the common example of
elelctric dipole is considered here.

Assuming the currents on the electric halfwave dipole to be
sinusoidal, the spherical harmonic expansion of the field
has been calculated in \cite{jackson1999classical}. The
corresponding coefficients in the expansion are
\begin{align*}{}
  a_E(1,0) &= \sqrt\frac{6}{\pi} \frac{I}{\lambda /2 }\\
  a_E(3,0) &= 49.5\times 10^{-3}\; a_E(1,0)\\
  a_E(5,0) &= 1.02\times 10^{-3}\; a_E(1,0)\\
\end{align*}

\begin{figure}[tbh]
  \centering
  \includegraphics[width=7cm]{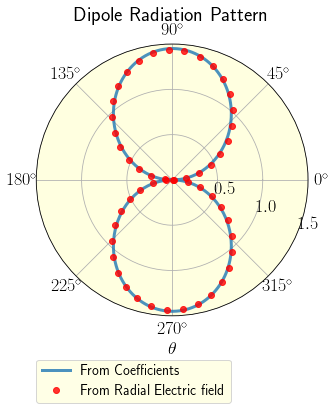}
  \caption{This is the comparision
    of the  far-field radiation for
    the dipole calculated directly
    from the coefficients and from
    the radial electric field
    expansion. They agree well. }
  \label{fig:Comparision}
\end{figure}

\begin{figure}[tbh]
  \centering
  \includegraphics[width=8.5cm]{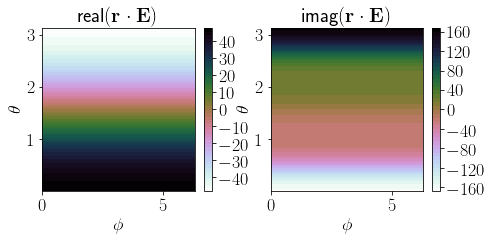}
  \caption{The real and imaginary parts of the "source"
    $\mathbf{r \centerdot E}$ on the $\lambda / 4$ sphere surrounding
    the halfwave dipole antenna.}
  \label{fig:rE_source}
\end{figure}

As this expansion has  electric multipoles coefficients
alone, the radial electric field on the sphere of radius $\lambda /4$
around the antenna can be estimated using
eq.~(\ref{eq:E-expanded}). The resulting "source"for the multipole
expansion in terms of the radial component of the electric field  is shown in Fig
\ref{fig:rE_source}. Using this  the far-fields
calculated using this method 
are shown in Fig \ref{fig:Comparision}, which correspond to
dipole radiation pattern in the far-field which has $\theta$-component
of the electric field alone.
This validates the radial-component being sufficient for finding the
far-field.

The fields of the magnetic dipole are dual to that of the electric
dipole which are related as

\[ \mathbf{E}_{\text{Magnetic Dipole}}\rightarrow -Z_0
  \mathbf{H}_{\text{Electric Dipole}}\]
  
\[ \mathbf{H}_{\text{Magnetic Dipole}}\rightarrow \frac{1}{Z_0}
  \mathbf{E}_{\text{Electric Dipole}}\]

The field expansion for the magnetic dipole is

  \[
    \boldsymbol{E}=-Z_0\sum_{l,m}\left[a_{M}(l,m)h_{l}\left(kr\right)\mathbf{X}_{lm}\right]
  \]
  \[
    \boldsymbol{H}=\frac{1}{Z_{0}}\sum_{l,m}\left[\cfrac{i}{k}a_{M}(l,m)\nabla\times\left(h_{l}\left(kr\right)\mathbf{X}_{lm}\right)\right]
  \]

  Hence the radial magnetic field will be same as that of the radial
  electric field shown in Fig \ref{fig:rE_source}. This "source" leads
  to far-field with $H_\theta$ and $E_\phi$ same as shown in Fig \ref{fig:Comparision}.
  Hence it is evident radial component being sufficient to represent
  the fields-outside the sphere enclosing the sources is also valid for the
  magnetic dipole.

\end{document}